\documentclass[aps,prd,twocolumn,preprintnumbers,superscriptaddress,nofootinbib]{revtex4-2}
\usepackage[dvipsnames]{xcolor}
\definecolor{red}{rgb}{0.9, 0,0}
\definecolor{cerulean}{rgb}{0., 0.42,0.9}
\definecolor{navy}{rgb}{0.05, 0.05,0.8}

\usepackage[colorlinks]{hyperref}
\hypersetup{
    colorlinks = true,
    citecolor  = red,
	linkcolor  = navy
}

\usepackage{mathtools}
\usepackage{amsmath}
\usepackage{amssymb}
\usepackage{amsfonts}
\usepackage{hyperref}
\usepackage{txfonts}
\usepackage{graphics}
\usepackage{endnotes}

\DeclareMathOperator{\sinc}{sinc}

\def\vec#1{\mathbf{#1}}
\newcommand{\unit}[1]{\vec{\hat{#1}}}

\begin{document}

\title{Proper Time Observables of General Gravitational Perturbations in Laser Interferometry-based Gravitational Wave Detectors}

\author{Vincent S. H. Lee}
\email{vincentszehimlee@berkeley.edu}
\affiliation{Walter Burke Institute for Theoretical Physics, California Institute of Technology, Pasadena, CA 91125, USA}
\affiliation{Department of Physics, University of California Berkeley, Berkeley, CA 94720, USA}
\affiliation{Department of Physics, University of California, San Diego, La Jolla, CA 92093, USA}
\author{Kathryn M. Zurek}
\email{kzurek@caltech.edu}
\affiliation{Walter Burke Institute for Theoretical Physics, California Institute of Technology, Pasadena, CA 91125, USA}

\preprint{CALT-TH-2024-025}

\begin{abstract}
We present an explicitly gauge-invariant observable of {\em any} general gravitational perturbation, $h_{\mu\nu}$ (\textit{not} necessarily due to gravitational waves (GWs)), in a laser interferometry-based GW detector, identifying the  signature as the proper time elapsed of the beamsplitter observer, between two events: when a photon passes through the beamsplitter, and when the same photon returns to the beamsplitter after traveling through the interferometer arm and reflecting off the far mirror. Our formalism applies to simple Michelson interferometers and can be generalized to more advanced setups. We demonstrate that the proper time observable for a plane GW is equivalent to the detector strain commonly used by the GW community, though now the common framework can be easily generalized for other types of signals, such as dark matter clumps or spacetime fluctuations from quantum gravity. We provide a simple recipe for computing the proper time observable for a general metric perturbation in linearized gravity and explicitly show that it is invariant under diffeomorphisms of the perturbation, as any physical observable should be.
\end{abstract}

\maketitle
\newpage

\section{Introduction}
%%%%%%%%%%%%%%%%%%%%%%%%%%%%%%%%%%%%%%%%%%%%%%%%%
%
%             Introduction
%
%%%%%%%%%%%%%%%%%%%%%%%%%%%%%%%%%%%%%%%%%%%%%%%%%%
The Laser Interferometer Gravitational-Wave Observatory (LIGO) has revolutionized astrophysics with its detection of gravitational waves (GWs), with the first observation of a binary black hole merger in 2015~\cite{LIGOScientific:2016aoc}.    Since then, approximately one hundred black hole mergers~\cite{KAGRA:2021vkt} and a few binary star mergers~\cite{LIGOScientific:2017vwq, LIGOScientific:2017ync, LIGOScientific:2020aai} have also been detected by LIGO and its European counterpart, Virgo. 
As a result of this success, several next-generation GW detectors are in development, including the Laser Interferometer Space Antenna (LISA)~\cite{LISA:2017pwj}, the Cosmic Explorer (CE)~\cite{Reitze:2019iox, Evans:2021gyd}, the Einstein Telescope (ET)~\cite{Punturo:2010zz, Hild:2010id, Sider:2022ntn}, and the Neutron Star Extreme Matter Observatory (NEMO)~\cite{Hall:2019xmm, Ackley:2020atn} (see Refs.~\cite{Aggarwal:2020olq, Ballmer:2022uxx} for reviews).

In addition to the known astrophysical sources, other types of physics could generate a signature in a laser interferometer.
For example, a transiting dark matter clump can induce signals in GW detectors through both gravitational and additional long-range interactions~\cite{Hall:2016usm}. Although this phenomenon has been explored in the literature~\cite{Seto:2004zu, Graham:2015ifn, Kawasaki:2018xak, Jaeckel:2020mqa, Baum:2022duc, Lee:2022tsw}, an explicitly gauge-invariant calculation using the metric perturbation due to a Newtonian potential was only recently performed in Ref.~\cite{Du:2023dhk}. Additionally, recent proposals suggest that spacetime fluctuations due to quantum gravity effects in causal diamonds could potentially be detectable by laser interferometers~\cite{Verlinde:2019xfb, Zurek:2020ukz}, with the signal spectrum derived in Refs.~\cite{Li:2022mvy, Bub:2023bfi} (see Ref.~\cite{Vermeulen:2024vgl} for an experimental proposal).

As a result of the broad range of physics that could be explored by a laser interferometer, it is worthwhile to clarify the physical observable.  In this way, the common framework developed for GW signatures ({\em e.g.},~\cite{Maggiore:2007ulw, DeLuca:2019ufz}) can be easily generalized for a broader range of sources of metric perturbations in laser interferometers.  In this paper, we identify the physics observable of \textit{any} gravitational perturbation in an interferometer system, such as simple Michelson interferometers~\cite{Michelson:1887zz}, to be the \textit{proper time} elapsed as measured by the \textit{beamsplitter} between the moments when a photon first passes through the beamsplitter and when the same photon returns. Denoting the beamsplitter's worldline as $x_B^{\mu}$, its change in proper time is given by the integral
\begin{equation} \label{eqn:Nambu-Goto}
	\Delta \tau_a = \int \sqrt{-g_{\mu\nu}dx_B^{\mu}dx_B^{\nu}} \, .
\end{equation}
Here the subscript $a$ labels the interferometer arms and runs from $1,2$. The interference pattern observed by collecting the laser from the beamsplitter to the detector is determined by the \textit{difference} of the proper time elapsed between the two arms, \textit{i.e.}
\begin{equation}\label{eqn:proper_time_difference}
	\Delta \tau = \Delta \tau_1 - \Delta \tau_2 \, ,
\end{equation}
and the strain is given by
\begin{equation}\label{eqn:strain_def}
	h = \frac{\Delta \tau}{2L} \, ,
\end{equation}
where $L$ is the length of the interferometer arms. 

The interference pattern observed when the lasers are recollected to the photon detector is determined by the relative phase shift between the two arms. We now show that this interference effect can be directly related to the proper time observable using the eikonal approximation of geometric optics. We first write the four-vector potential of a plane electromagnetic wave in curved spacetime as
\begin{equation}\label{eqn:eikonal_approximation}
	A^{\mu}(x) = a^{\mu}(x)e^{i\varphi(x)} \, ,
\end{equation}
where $\varphi(x)$ and $a^{\mu}(x)$ are the phase and amplitude of the electromagnetic wave, respectively. Let $\lambda$ be an affine parameter of the photon worldline and define $k_{\mu}\equiv \partial_{\mu}\varphi$ to be its wave vector. In the geometric optics limit, we consider $\varphi(x)$ to be rapidly oscillating at length scale $\lambdabar$ and $a^{\mu}(x)$ to be slow-varying at a length scale $L_a$. To the lowest order in $(\lambdabar/L_a)$, using the Maxwell equation, $\nabla_{\mu}A^{\mu}(x)=0$, where $\nabla_{\mu}$ is the covariant derivative with respect to $g_{\mu\nu}$, one can derive (\textit{cf.} Eq.~(1.187) in Ref.~\cite{Maggiore:2007ulw})
\begin{equation}\label{eqn:eikonal_equation}
	g_{\mu\nu}k^{\mu}k^{\nu}=0 \, ,
\end{equation}
which is known as the \textit{eikonal equation}. On the other hand, the photon worldline is given by the null geodesic equation, $g_{\mu\nu}(dx^{\mu}/d\lambda)(dx^{\nu}/d\lambda)=0$. Combined with Eq.~\eqref{eqn:eikonal_equation}, this shows that $k^{\mu}\propto dx^{\nu}/d\lambda$ along the geodesic (\textit{i.e.} $k^{\mu}$ is a tangent vector of the geodesic). This implies the phase $\varphi$ is, in fact, a constant along the photon worldline, 
\begin{equation}\label{eqn:constant_phase}
	\frac{d}{d\lambda}\varphi = \partial_{\mu}\varphi \left(\frac{dx^{\mu}}{d\lambda}\right) \propto k_{\mu}k^{\mu} = 0 \, .
\end{equation}
If a particular photon arrives at the beamsplitter at proper time (as measured by the beamsplitter) $\tau$, then its phase is given by $\varphi = \omega_{\mathrm{laser}}\tau$, where $\omega_{\mathrm{laser}} = 2\pi / \lambdabar$ is the laser frequency. While $\omega_{\mathrm{laser}}$ can be changed by the metric fluctuation, to the lowest order in $(\lambdabar/L_a)$, the phase $\varphi$ is still conserved as the photon propagates through the curved spacetime. Thus, we can neglect any change in $\omega_{\mathrm{laser}}$ following the eikonal approximation (see also Ref.~\cite{Domcke:2024abs}).

We now consider a continuous stream of photons arriving at the beamsplitter from the laser source. Two particular photons pass through the beamsplitter at two (possibly different) proper times, $\tau_1$ and $\tau_2$ (as measured by the beamsplitter), along the two interferometer arms and recombine at the beamsplitter at the same proper time, $\tau_f$, such that the proper time shifts are computed by $\Delta\tau_{1,2}\equiv\tau_f-\tau_{1,2}$. Each photon admits the phase $A_1^{\mu}\propto\exp(i\omega_{\mathrm{laser}}\tau_1)$ and $A_2^{\mu}\propto\exp(i\omega_{\mathrm{laser}}\tau_2)$, respectively, which are conserved along the wavefront (up to any discontinuous phase shifts due to the mirrors and the beamsplitter, which are irrelevant as they can be reabsorbed into the definition of the unperturbed length of the cavities, which are fixed experimentally) until they recombine at the beamsplitter. Their superposition thus admits a relative phase of
\begin{align}\label{eqn:relative_phase}
	\Delta\phi=\omega_{\mathrm{laser}}(\tau_2-\tau_1)&=\omega_{\mathrm{laser}}[(\tau_f-\tau_1)-(\tau_f-\tau_2)] \nonumber \\
	&=\omega_{\mathrm{laser}}\Delta \tau \, ,
\end{align}
from Eq.~\eqref{eqn:proper_time_difference}. Since the proper time as defined in Eq.~\eqref{eqn:Nambu-Goto} is a scalar, it is manifestly invariant under diffeomorphisms on the metric perturbation. It might be surprising at first that the physical observable is a time shift quantity as measured by the beamsplitter, not the photons.\footnote{One might be tempted to attribute the physical interferometer observable to either the change in \textit{proper distance} between the beamsplitter and the far mirror, or the change in \textit{proper time} for the light propagating between them. However, neither description is entirely satisfying. Proper distances are only defined between two spacelike-separated events, which cannot generally be measured by null geodesics unless the proper distance does not change substantially within a photon roundtrip time (which holds for long-wavelength GWs but is not true in general). On the other hand, proper times are ill-defined for photons as they travel in null geodesics (\textit{i.e.} $ds^2=0$).} Indeed, if we consider a ``triangular path" formed by joining the beamsplitter worldline with the outgoing and incoming photon worldlines and evaluate the loop integral $\Delta \tau_a = \oint \sqrt{-g_{\mu\nu} dx^\mu dx^\nu}$ over the three paths, we find that the integral vanishes for the two photon worldlines since photons travel along null geodesics. Thus, the loop integral reduces to a single line integral over the beamsplitter worldline, as in Eq.~\eqref{eqn:Nambu-Goto}. However, any deviation in the photon trajectories will alter the proper time measured by the beamsplitter by changing the integration limits in Eq.~\eqref{eqn:Nambu-Goto}. Specifically, for GW signals, as we will later show, the proper time observable is equivalent to the commonly quoted detector strain, which is well known to be gauge-invariant by calculations in both the transverse-traceless (TT) gauge and the proper detector frame~\cite{Maggiore:2007ulw}.\footnote{A calculation of GW measurements with laser interferometers in a general gauge, assuming slowly varying scalar and vector perturbations along a photon roundtrip, is recently performed in Ref.~\cite{DeLuca:2019ufz}.} Furthermore, perturbations on the photons and the beamsplitter (as well as the far mirror) can be shuffled into each other by a generic gauge transformation, while the sum of all effects remains gauge-invariant. We will explicitly verify these statements in later sections.

The goal of this paper is to fully develop the idea of proper time as a physical observable, focused on the case of a simple Michelson interferometer. We provide a simple recipe for computing the observable given a generic metric perturbation in any gauge and show that, for any metric perturbation, the beamsplitter proper time elapsed can be written as a sum of Doppler, Shapiro, and Einstein delays. This latter result is consistent with previous work~\cite{Rakhmanov:2004eh, Li:2022mvy}, though now we are able to show explicitly that this sum is the proper time elapse of the beamsplitter observer. This recipe can now be applied to any metric perturbation beyond the usual GW treatment. Readers interested in computing this gauge-invariant observable can directly refer to Eqs.~\eqref{eqn:proper_time_difference}--\eqref{eqn:strain_def} and Eqs.~\eqref{eqn:unperturbed_worldlines}--\eqref{eqn:dse}. Alternatively, readers can first perform a gauge transformation of the given metric into a particularly convenient gauge, introduced in Eq.~\eqref{eqn:synchronous_transformation}, which significantly simplifies the calculation by pushing all observable effects from the metric to the perturbation on the photon worldlines, analogous to the TT gauge for GWs. In Eq.~\eqref{eqn:Doppler_gauge}, Eq.~\eqref{eqn:Shapiro_gauge_2}, and Eq.~\eqref{eqn:Einstein_gauge}, we explicitly demonstrate that the proper time observable is gauge-invariant by performing a generic gauge transformation on the metric perturbation.

We note that our treatment differs from the existing literature. Gauge invariance in GW detection has been widely discussed in the GW community. For instance, Ref.~\cite{Rakhmanov:2004eh} showed that the interferometer response to GWs in the TT gauge and the proper detector frame agrees when summing over three distinct contributions that are not individually gauge-invariant. In addition, Ref.~\cite{Finn:2008np} identified errors in existing derivations of the interferometer response and provided a corrected calculation of the GW interferometer response in the TT gauge by computing the proper time shift, taking into account perturbations on the photon and beamsplitter/mirror worldlines. While these works have resolved gauge-invariance issues for GWs in two popular gauges, the questions of gauge invariance and its relation to the proper time observable in other GW gauges or for other non-GW metric perturbations due to beyond-the-standard-model (BSM) physics, remain. The goal of this paper is to address this issue.

\section{Proper time observable from generic metric perturbations}
\begin{figure*}
	\includegraphics[width=1.\textwidth]{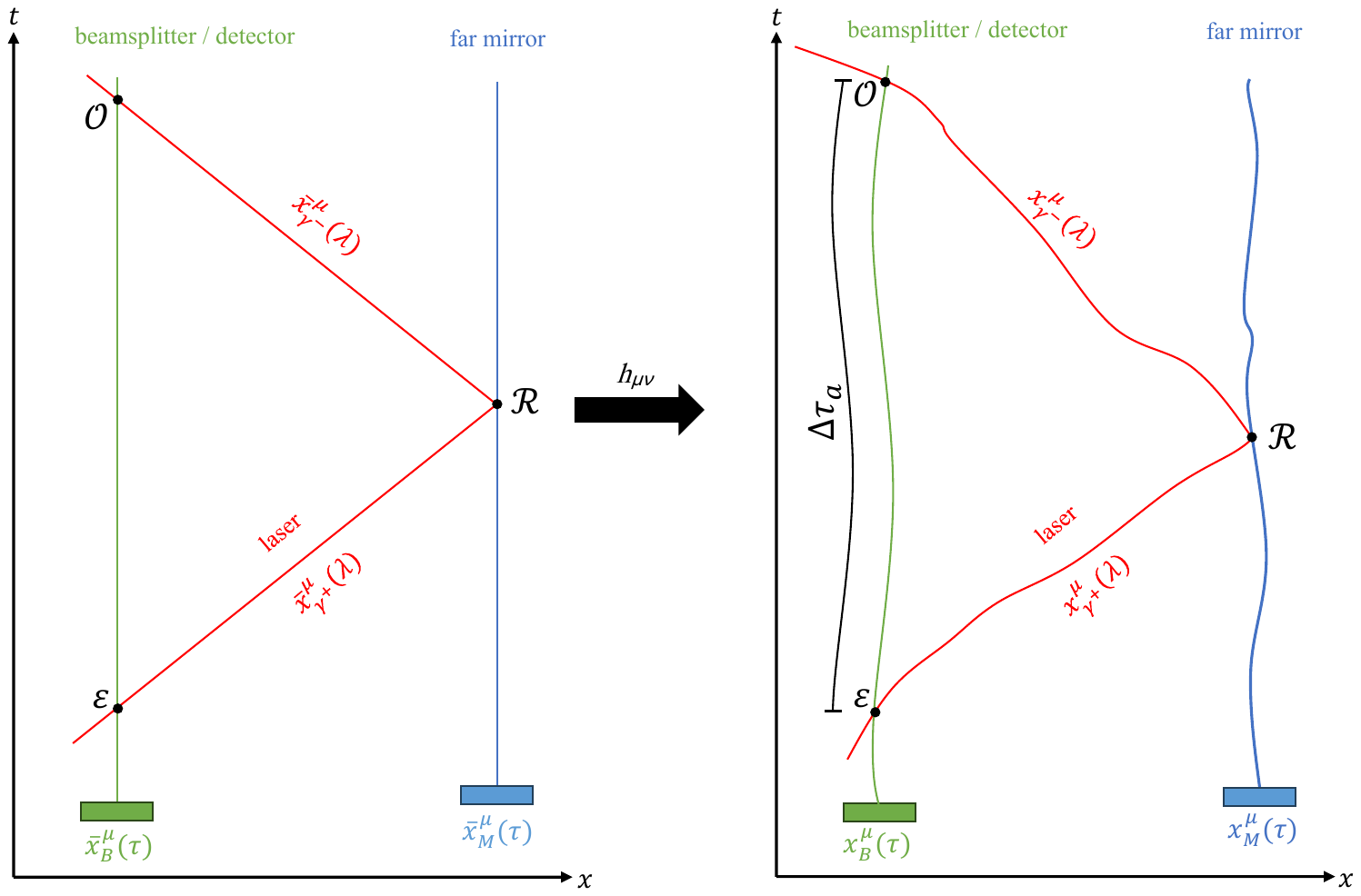} 
	\caption{Schematic spacetime diagram of a single arm in a GW detector. The left/right diagram corresponds to the unperturbed/perturbed system. The worldlines of the beamsplitter, the far mirror, and the photon are represented by green, blue, and red lines, and denoted by $x^{\mu}_B(\tau)$, $x^{\mu}_M(\tau)$, and $x^{\mu}_{\gamma^{\pm}}(\lambda)$, respectively. The physical events, emission $(\mathcal{E})$, reflection $(\mathcal{R})$, and observation $(\mathcal{O})$ are defined by the intersections of worldlines (see description of each event in the main text). The proper time observable for each arm, $\Delta \tau_a$, is defined to be the proper time elapsed at the beamsplitter between $\mathcal{E}$ and $\mathcal{O}$. Quantities defined in the unperturbed system are denoted by an overbar.}\label{fig:setup}
\end{figure*}

%%%%%%%%%%%%%%%%%%%%%%%%%%%%%%%%%%%%%%%%%%%%%%%%%
%
%       Proper time observable from generic metric perturbations
%
%%%%%%%%%%%%%%%%%%%%%%%%%%%%%%%%%%%%%%%%%%%%%%%%%%
A laser interferometer consists of two arms linked by a beamsplitter. We schematically depict a single arm of a laser interferometer in Fig.~\ref{fig:setup}. For a single measurement, there are three notable physical events:
\begin{enumerate}
	\item \textbf{Emission} ($\mathcal{E}$): The laser reaches the beamsplitter and splits into two beams, each traversing along separate arms.
	\item \textbf{Reflection} ($\mathcal{R}$): The beams encounter the mirrors at the far ends, and reflect back toward the beamsplitter.
	\item \textbf{Observation} ($\mathcal{O}$): The beams return to the beamsplitter, and finally reconverge at the detection port. The intensity of the laser is then measured to determine the phase difference between the two laser beams.%
\end{enumerate}
Suppose the metric is perturbed as $g_{\mu\nu} = \eta_{\mu\nu} + h_{\mu\nu}$, where $\eta_{\mu\nu}=\mathrm{diag}(-1,+1,+1,+1)$ is the metric of the flat Minkowski spacetime in the mostly positive convention. When the perturbation is small, we apply the linearized gravity limit, and only terms with the leading order in $h_{\mu\nu}$ are retained. 

A general gauge transformation on the metric perturbation is defined as
\begin{equation}\label{eqn:gauge_transformation}
	h_{\mu\nu} \to h'_{\mu\nu} = h_{\mu\nu}-\partial_{\mu}\xi_{\nu}-\partial_{\nu}\xi_{\mu} \, ,
\end{equation}
for a general vector field $\xi^{\mu}$. The physical observable must be a quantity that is invariant under a transformation as defined in Eq.~\eqref{eqn:gauge_transformation}. One such quantity is given by the proper time elapsed, as measured by the photon detector, between photon emission at the beamsplitter ($\mathcal{E}$) and when the photon is reflected back to the beamsplitter and sent to the detection port ($\mathcal{O}$). Here, we assume that the detection port and the mirrors close to the beamsplitter are positioned closely to the beamsplitter itself (in comparison to the interferometer arm length), and thus can all be regarded as the same point in spacetime. 

We first consider a single arm in the interferometer oriented toward the unit vector $\unit{n}$. The proper time elapsed, as measured by the detector, is given by
\begin{align} \label{eqn:proper_action}
	\Delta \tau_a = \int_{\tau_{\mathcal{E}}}^{\tau_{\mathcal{O}}} d\tau = \tau_{\mathcal{O}}-\tau_{\mathcal{E}}  \, ,
\end{align}
where $\tau_{\mathcal{E}}$ and $\tau_{\mathcal{O}}$ denote the detector proper times when the photon is emitted from the beamsplitter and observed by the detector, respectively. In our setup, we assume that $\tau_{\mathcal{E}}$ is independent of the metric perturbation since it simply marks the starting (proper) time of a measurement. In this case, $\tau_{\mathcal{O}}$ will change under a metric perturbation, determined by the intersection of worldlines between the detector and the photon. We note that one could instead choose to fix $\tau_{\mathcal{O}}$ as independent of the metric perturbation and compute the corresponding change in $\tau_{\mathcal{E}}$, which would still lead to the same result for the proper time shift in Eq.~\eqref{eqn:proper_action}, following the steps outlined in the remainder of the paper.

In a general perturbed metric, the (3+1)-dimensional worldlines do not necessarily intersect with each other. Nevertheless, in a realistic experimental setup, the beamsplitter and the mirrors have finite extents, which are usually neglected since they are much shorter than the interferometer arm length. In the $h_{\mu\nu}\ll 1$ limit, the photons should be considered as coinciding with the detector / mirrors as long as the projections of their worldlines on the $\unit{n}$ direction overlap. Denoting the outgoing and incoming photon worldlines by $x^{\mu}_{\gamma^+}(\lambda)$ and $x^{\mu}_{\gamma^-}(\lambda)$, respectively with $\lambda$ being an affine parameter, and the detector and far mirror worldlines by $x^{\mu}_B(\tau)$ and $x^{\mu}_M(\tau)$, respectively, these intersections are then obtained by solving the system of equations 
\begin{align}\label{eqn:system_of_equation}
	\mathcal{E}:&\begin{dcases*}
		x^0_B(\tau_{\mathcal{E}}) = x^{0}_{\gamma^+}(\lambda_{\mathcal{E}})  \\
		\unit{n}\cdot \vec{x}_B(\tau_{\mathcal{E}}) = \unit{n}\cdot\vec{x}_{\gamma^+}(\lambda_{\mathcal{E}})  \\
	\end{dcases*} \nonumber \\
	\mathcal{R}:&\begin{dcases*}
	x^0_M(\tau_{\mathcal{R}}) = x^{0}_{\gamma^+}(\lambda_{\mathcal{R}}) = x^{0}_{\gamma^-}(\lambda_{\mathcal{R}}) \\
	\unit{n}\cdot \vec{x}_M(\tau_{\mathcal{R}}) = \unit{n}\cdot\vec{x}_{\gamma^+}(\lambda_{\mathcal{R}}) = \unit{n}\cdot\vec{x}_{\gamma^-}(\lambda_{\mathcal{R}})\\
	\end{dcases*} \nonumber \\
	\mathcal{O}:&\begin{dcases*}
		x^0_B(\tau_{\mathcal{O}}) = x^{0}_{\gamma^-}(\lambda_{\mathcal{O}})  \\
	\unit{n}\cdot \vec{x}_B(\tau_{\mathcal{O}}) = \unit{n}\cdot\vec{x}_{\gamma^-}(\lambda_{\mathcal{O}}) \, ,
	\end{dcases*} \, ,
\end{align}
where $\tau_{\mathcal{R}}$ is the mirror proper time when it reflects the photon, and $\lambda_{\mathcal{E}}$, $\lambda_{\mathcal{R}}$, and $\lambda_{\mathcal{O}}$ are the photon affine parameters when it is emitted, reflected and observed, respectively. In general, Eq.~\eqref{eqn:system_of_equation} needs to be solved by finding the intersections of geodesics under a general metric. In linearized gravity, however, we can solve for worldline intersections order by order. Denoting unperturbed worldline trajectories and time quantities by an overbar and perturbation by $\delta$, we can expand a generic timelike geodesic in the leading order of $h$ as $x_{B,M}^{\mu}(\tau)=\bar{x}_{B,M}^{\mu}(\bar{\tau})+\delta \tau\,\bar{v}_{B,M}^{\mu}(\bar{\tau})+\delta x_{B,M}^{\mu}(\bar{\tau})$ with $\tau=\bar{\tau}+\delta \tau$, $\bar{v}_{B,M}^{\mu}(\tau)=(d/d\tau)\bar{x}_{B,M}^{\mu}(\tau)$, and analogous equations for the null geodesics. Hence, Eq.~\eqref{eqn:system_of_equation} becomes
\begin{align}\label{eqn:system_of_equation_solved}
	\mathcal{E}:&\begin{dcases*}
		\delta x^0_B(\tau_{\mathcal{E}}) = \delta x^0_{\gamma^+}(\lambda_{\mathcal{E}})  \\
		\unit{n}\cdot\delta \vec{x}_B(\tau_{\mathcal{E}}) = \unit{n}\cdot\delta \vec{x}_{\gamma^+}(\lambda_{\mathcal{E}}) \\
	\end{dcases*} \nonumber \\
	\mathcal{R}:&\begin{dcases*}
		\delta x^0_M(\bar{\tau}_{\mathcal{R}})+\delta\tau_{\mathcal{R}} = \delta x^0_{\gamma^+}(\bar{\lambda}_{\mathcal{R}})+\delta\lambda_{\mathcal{R}} = \delta x^0_{\gamma^-}(\bar{\lambda}_{\mathcal{R}})+\delta\lambda_{\mathcal{R}} \\
		\unit{n}\cdot\delta \vec{x}_M(\bar{\tau}_{\mathcal{R}}) = \unit{n}\cdot\delta \vec{x}_{\gamma^+}(\bar{\lambda}_{\mathcal{R}})+\delta \lambda_{\mathcal{R}} = \unit{n}\cdot\delta \vec{x}_{\gamma^-}(\bar{\lambda}_{\mathcal{R}})-\delta \lambda_{\mathcal{R}} \\
	\end{dcases*} \nonumber \\
	\mathcal{O}:&\begin{dcases*}
		\delta x^0_B(\bar{\tau}_{\mathcal{O}})+\delta \tau_{\mathcal{O}} = \delta x^0_{\gamma^-}(\bar{\lambda}_{\mathcal{O}})+\delta \lambda_{\mathcal{O}}  \\
		\unit{n}\cdot\delta \vec{x}_B(\bar{\tau}_{\mathcal{O}}) = \unit{n}\cdot\delta \vec{x}_{\gamma^-}(\bar{\lambda}_{\mathcal{O}})-\delta \lambda_{\mathcal{O}}
	\end{dcases*} \, ,
\end{align}
where we used the expressions for the unperturbed worldlines %
\begin{align}\label{eqn:unperturbed_worldlines}
	\bar{x}_B^{\mu}(\tau) &= (\tau,\bar{\vec{x}}_B) \nonumber \\
	\bar{x}_M^{\mu}(\tau) &= (\tau,\bar{\vec{x}}_M) \nonumber \\
	\bar{x}^{\mu}_{\gamma^+}(\lambda) &= (\lambda,(\lambda-\lambda_{\mathcal{E}})\unit{n}+\bar{\vec{x}}_B) \nonumber \\
	\bar{x}^{\mu}_{\gamma^-}(\lambda) &= (\lambda,-(\lambda-\bar{\lambda}_{\mathcal{O}})\unit{n}+\bar{\vec{x}}_B) \, .
\end{align}
We note that the asymmetry between the emission equations (first two) and the observation equations (last two) in Eq.~\eqref{eqn:system_of_equation_solved} originates from the fact that $\delta \tau_{\mathcal{E}}=0$ (\textit{i.e.} the emission proper time is unperturbed) while $\delta \tau_{\mathcal{O}}\neq 0$. %

The perturbed worldlines of the detector and the far mirror are given by solving the geodesic equation, $d^2x_{B,M}^{\rho}/d\tau^2+\Gamma^{\rho}_{\mu\nu}(dx_{B,M}^{\mu}/d\tau)(dx_{B,M}^{\nu}/d\tau)=0$, which in linearized gravity gives
\begin{equation}\label{eqn:timelike_geodesic}
	\frac{d^2}{d\tau^2}\delta x^i_{B,M}(\tau) = \eta^{ij}\left(-\partial_0h_{0j}+\frac{1}{2}\partial_jh_{00}\right)\bigg|_{\bar{x}^{\mu}_{B,M}(\tau)}  \, .
\end{equation}
The time component of the worldlines are determined by $ds^2=g_{\mu\nu}dx^{\mu}_{B,M}dx^{\nu}_{B,M}=-1$ for timelike geodesics
\begin{equation}\label{eqn:timelike_geodesic_time}
	\frac{d}{d\tau}\delta x^0_{B,M}(\tau) = \frac{1}{2}h_{00}\bigg|_{\bar{x}^{\mu}_{B,M}(\tau)}  \, .
\end{equation}
In Eqs.~\eqref{eqn:timelike_geodesic}--\eqref{eqn:timelike_geodesic_time}, the metric derivatives on the RHS are evaluated at the unperturbed detector and mirror worldlines, $\bar{x}^{\mu}_B(\tau)$ and $\bar{x}^{\mu}_M(\tau)$, as defined in Eq.~\eqref{eqn:unperturbed_worldlines}, respectively. These differential equations are solved by integrating the RHS of Eqs.~\eqref{eqn:timelike_geodesic}--\eqref{eqn:timelike_geodesic_time} in $\tau$. The photon geodesics are solved by setting $ds^2=g_{\mu\nu}dx_{\gamma^{\pm}}^{\mu}dx_{\gamma^{\pm}}^{\nu}=0$, which becomes
\begin{equation}\label{eqn:null_geodesic}
	n^{\pm}_{\mu}\delta v^{\mu}_{\gamma^{\pm}}(\lambda) = -\frac{1}{2}\left[h_{00}\pm 2n^ih_{0i}+n^in^jh_{ij}\right]\bigg|_{\bar{x}^\mu_{\bar{\gamma}^{\pm}}(\lambda)} \, ,
\end{equation}
where we defined $v^{\mu}_{\gamma^{\pm}}(\lambda)=(d/d\lambda)x^{\mu}_{\gamma^{\pm}}(\lambda)$, $n_{\pm}^{\mu}=(1,\pm n^i)$, and the perturbed metric in the RHS are evaluated at the unperturbed photon trajectories, $\bar{x}^{\mu}_{\gamma^{\pm}}(\lambda)$, as defined in Eq.~\eqref{eqn:unperturbed_worldlines}, respectively. 

Finally, we rearrange Eq.~\eqref{eqn:system_of_equation_solved} to solve for $\Delta \tau_a = \tau_{\mathcal{O}} - \tau_{\mathcal{E}} = 2L + \delta\tau_{\mathcal{O}}$. Specifically, we take each individual equation in Eq.~\eqref{eqn:system_of_equation_solved}, add and subtract them in such a way that terms involving $\delta x^0_M(\bar{\tau}_R)$, $\delta \lambda_{\mathcal{R}}$, and $\delta \lambda_{\mathcal{O}}$ cancel out, and the temporal and spatial components of the deviation in photon worldlines combine into $n^{\pm}_{\mu} \delta x^{\mu}_{\gamma^{\pm}}$. After some algebra, we find that Eq.~\eqref{eqn:proper_action} can be written in a compact form
\begin{equation}\label{eqn:proper_time_compact}
	\Delta \tau_a = 2L+\Delta \tau_a^{(\mathrm{Doppler})} + \Delta \tau_a^{(\mathrm{Shapiro})} +\Delta \tau^{(\mathrm{Einstein})}   \, ,
\end{equation}
where we have written the shift in proper time elapsed as a sum of three individual contributions, defined as
\begin{align}\label{eqn:dse}
	\Delta \tau^{(\mathrm{Doppler})}_a &= -\unit{n}\cdot\left[\delta \vec{x}_B(\tau_{\mathcal{E}})-2\delta\vec{x}_M(\tau_{\mathcal{E}}+L)+\delta \vec{x}_B(\tau_{\mathcal{E}}+2L)\right] \nonumber \\
	\Delta \tau^{(\mathrm{Shapiro})}_a &= -\int_{\lambda_{\mathcal{E}}}^{\lambda_{\mathcal{E}}+L}d\lambda\,n^{+}_{\mu}\delta v_{\gamma^+}^{\mu}(\lambda) - \int_{\lambda_{\mathcal{E}}+L}^{\lambda_{\mathcal{E}}+2L}d\lambda\,n^-_{\mu}\delta v_{\gamma^-}^{\mu}(\lambda) \nonumber \\
	\Delta \tau^{(\mathrm{Einstein})} &= -\int_{\tau_{\mathcal{E}}}^{\tau_{\mathcal{E}}+2L}d\tau\,\left[\frac{d}{d\tau}\delta x^0_B(\tau)\right] \, ,
\end{align}
where $\lambda_{\mathcal{E}}=\tau_{\mathcal{E}}$, and the expressions for the worldlines are given in Eqs.~\eqref{eqn:unperturbed_worldlines}--\eqref{eqn:null_geodesic}. Here we used $\bar{\tau}_{\mathcal{O}}=\tau_{\mathcal{E}}+2L$ and $\bar{\tau}_{\mathcal{R}}=\tau_{\mathcal{E}}+L$. The difference in proper time elapsed between two arms is then given by Eq.~\eqref{eqn:proper_time_difference}, Eq.~\eqref{eqn:proper_time_compact}, and Eq.~\eqref{eqn:dse}. An analogous decomposition has also been identified in Ref.~\cite{Rakhmanov:2004eh} for the special case of interferometer response to GWs in the proper detector frame, in which the sum is shown to be translationally invariant, although our treatment now applies more generally to any gauge of a general metric perturbation, not necessarily GWs. 

We briefly describe the physical meanings of the Doppler, Shapiro, and Einstein terms defined in Eq.~\eqref{eqn:dse}. The Doppler term corresponds to the motion of the detector and the far mirror, akin to the Doppler effect, where the motion of the emitter or observer of a wave affects the apparent frequency. The Shapiro term accounts for the integrated shifts in velocity in the photon trajectories, similar to a Shapiro delay for light waves propagating under the influence of a gravitational field. Lastly, the Einstein term corresponds to the time dilation of the clock at the detector.

Since the proper time elapsed defined in Eq.~\eqref{eqn:proper_action} is a scalar, it is by construction a gauge-invariant quantity. We will explicitly verify this by performing a general gauge transformation on the metric as Eq.~\eqref{eqn:gauge_transformation} and computing the change in each term in Eq.~\eqref{eqn:proper_time_compact}. One finds in Eq.~\eqref{eqn:Doppler_gauge}, Eq.~\eqref{eqn:Shapiro_gauge_2} and Eq.~\eqref{eqn:Einstein_gauge} that while individual contributions might be altered, the sum of all three contributions remain unchanged. This highlights an important observation: although one could attribute physical meanings to the individual components in Eq.~\eqref{eqn:proper_time_compact}, such as proper motion and time delay, they are ultimately coordinate-dependent quantities, and by themselves do not constitute a true physical observable as measured by a realistic experiment. 

We conclude this section with several remarks. First, observe that the Einstein term does not depend on the arm orientation $\unit{n}$, and hence is completely canceled when the proper time difference across the two arms is computed. Second, the time variable of this measurement can be identified as the final time of the light-pulse sequence. Therefore, the power spectrum of the signal is obtained by Fourier transforming the proper time elapsed with respect to $\tau_{\mathcal{O}}=\tau_{\mathcal{E}}+2L+\mathcal{O}(h)$, the time of observation, \textit{i.e.}~\cite{Moore:2014lga}
\begin{align}\label{eqn:strain}
	\tilde{h}(f) = \frac{1}{2L}\int_{-\infty}^{\infty}d\tau_{\mathcal{E}}\,e^{-2\pi if(\tau_{\mathcal{E}}+2L)}\Delta \tau  \, .
\end{align}

Finally, as an example, we compute the proper time observable as defined above for a classical GW. Assuming that the wave passes through the interferometer at $t=0$ in the $\unit{z}$ direction while the two interferometer arms are oriented in the $\unit{x}$ and $\unit{y}$ directions and the beamsplitter set at the origin, the metric perturbation, in the TT gauge, can be written as $h^{TT}_{xx}=-h^{TT}_{yy}=h_+\cos(\omega_{\mathrm{GW}} t-k_{\mathrm{GW}}z)$ and $h^{TT}_{xy}=h^{TT}_{yx}=h_{\times}\cos(\omega_{\mathrm{GW}} t-k_{\mathrm{GW}}z)$, where $h_+$ and $h_{\times}$ are the plus-polarization and cross-polarization strains, and $\omega_{\mathrm{GW}}$ and $k_{\mathrm{GW}}$ are the angular frequency and wavenumber of the GW. It is immediately clear from Eqs.~\eqref{eqn:timelike_geodesic}--\eqref{eqn:null_geodesic} that the only term in Eq.~\eqref{eqn:dse} that contributes to the proper time shift is the Shapiro term, which evaluates to (using Eqs.~\eqref{eqn:null_geodesic}--\eqref{eqn:dse} and Eq.~\eqref{eqn:proper_time_difference})
\begin{equation}\label{eqn:strain_GW}
	\Delta\tau^{(\mathrm{GW})} = 2Lh_+\sinc\left(\omega_{\mathrm{GW}} L\right)\cos\left(\omega_{\mathrm{GW}}(\tau_{\mathcal{E}}+L)\right)  \, .
\end{equation}
The strain as derived from the proper time observable by putting Eq.~\eqref{eqn:strain_GW} into Eq.~\eqref{eqn:strain_def} is then given by the familiar relation, $h^{(\mathrm{GW})}(t)\approx h_+\cos(\omega_{\mathrm{GW}}t)$~\cite{Cahillane:2022pqm} in the long-wavelength limit, $\omega_{\mathrm{GW}} L\ll1$. We have thus shown that the proper time observable is equivalent to the detector strain for GWs. Consequently, when discussing GW signatures in the TT gauge, one can focus solely on the photon time delay across a round-trip as a simplified definition of the observable. 

As a sanity check, we repeat the above calculation in the proper detector frame. Using results from Ref.~\cite{Marzlin:1994ia}, the only nonvanishing and relevant component is given by $h^{PD}_{00}\Big|_{z=0}=-(1/2)\omega_{\mathrm{GW}}^2\left[h_{+}(x^2-y^2)+2h_{\times}xy\right]\cos(\omega_{\mathrm{GW}}t)$~\cite{Berlin:2021txa}. One finds that both Doppler and Shapiro terms contribute, with (\textit{cf. }Eqs.~\eqref{eqn:null_geodesic}--\eqref{eqn:dse} and Eq.~\eqref{eqn:proper_time_difference})
\begin{align}\label{eqn:GW_PD}
	\Delta\tau^{\mathrm{(GW,Doppler)}}&=2Lh_+\cos(\omega_{\mathrm{GW}}(\tau_{\mathcal{E}}+L)) \nonumber \\
	\Delta\tau^{\mathrm{(GW,Shapiro)}}&=-2Lh_+\cos(\omega_{\mathrm{GW}}(\tau_{\mathcal{E}}+L))\left[1-\sinc(\omega_{\mathrm{GW}}L)\right] \, ,
\end{align}
which reproduces the TT-gauge result in Eq.~\eqref{eqn:strain_GW}. We note that in the proper detector frame, the mirror motion (\textit{i.e.} Doppler term) is the dominant contribution if $\omega_{\mathrm{GW}}L\ll 1$, in agreement with the literature~\cite{Maggiore:2007ulw}.

\section{A convenient gauge choice: the synchronous gauge}
%%%%%%%%%%%%%%%%%%%%%%%%%%%%%%%%%%%%%%%%%%%%%%%%%
%
%      A convenient gauge choice - the synchronous gauge
%
%%%%%%%%%%%%%%%%%%%%%%%%%%%%%%%%%%%%%%%%%%%%%%%%%%
We have provided a formalism of computing the interferometer observable of a generic metric perturbation. While the observable is independent of the gauge choice, it is often most convenient to work in the \textit{synchronous} gauge, defined as $h_{00}=h_{0i}=0$~\cite{Lifshitz:1945du}. Similar to the TT gauge for GWs, the only non-vanishing contribution to the proper time measurement in Eq.~\eqref{eqn:dse} is the Shapiro term, and hence all physical effects are put into the deviation in photon geodesics, and the needs for solving geodesic equations are evaded. A metric perturbation, $h_{\mu\nu}$, can be brought into the synchronous gauge using the gauge transformation in Eq.~\eqref{eqn:gauge_transformation} with the vector field defined as indefinite time integrals of $h_{\mu\nu}$,
\begin{align}\label{eqn:synchronous_transformation}
	\xi_0(t,\vec{x}) &= \frac{1}{2}\int^{t}dt'\,h_{00}(t',\vec{x}) \nonumber \\
	\xi_i(t,\vec{x}) &= \int^{t}dt'\,h_{0i}(t',\vec{x}) - \frac{1}{2}\partial_i\int^tdt'\int^{t'}dt''\,h_{00}(t'',\vec{x}) \, ,
\end{align}
which are effectively solutions to setting Eq.~\eqref{eqn:timelike_geodesic} to zero, such that massive bodies appear at rest in the new coordinates. Here we note that there are residual degrees of freedom in the synchronous gauge.

\section{Explicit demonstration of gauge invariance}
%%%%%%%%%%%%%%%%%%%%%%%%%%%%%%%%%%%%%%%%%%%%%%%%%
%
%       Explicit demonstration of gauge invariance
%
%%%%%%%%%%%%%%%%%%%%%%%%%%%%%%%%%%%%%%%%%%%%%%%%%%
The proper time measured by the beamsplitter as defined in Eq.~\eqref{eqn:Nambu-Goto} is a scalar, and is thus manifestly gauge-invariant. In this section, we reinforce this point by explicitly performing a gauge transformation on the metric, as defined in Eq.~\eqref{eqn:gauge_transformation}, and demonstrating that the proper time computed from Eqs.~\eqref{eqn:timelike_geodesic}--\eqref{eqn:dse} is unaltered. Here we note that a similar verification for a more restricted class of metric perturbations has been performed in App. B of Ref.~\cite{Li:2022mvy}.

We start with the Doppler term. To solve the timelike geodesic equations of the beamsplitter and the mirrors in Eq.~\eqref{eqn:timelike_geodesic}, we must first specify their initial conditions, $\delta \vec{x}_{B,M}(\tau_0)$ and $\delta \vec{v}_{B,M}(\tau_0)$, where we defined $\delta \vec{v}_{B,M}(\tau)=(d/d\tau)\delta \vec{x}_{B,M}(\tau)$, and $\tau_0$ to be some initial proper time before the measurements are made. Applying the gauge transformation as defined in Eq.~\eqref{eqn:gauge_transformation} will also shift the initial conditions as $\delta x_{B,M}^i(\tau_0)\to \delta x_{B,M}^i(\tau_0)+\xi^i(\bar{x}^{\mu}_{B,M}(\tau_0))$ and $\delta v_{B,M}^i(\tau_0)\to \delta v_{B,M}^i(\tau_0)+\partial_0\xi^i(\bar{x}^{\mu}_{B,M}(\tau_0))$. Integrating Eq.~\eqref{eqn:timelike_geodesic} in $\tau$ twice, we find that the Doppler term in Eq.~\eqref{eqn:dse} in the new gauge is  
\begin{align}\label{eqn:Doppler_gauge}
	\Delta \tau^{(\mathrm{Doppler})\prime} &= 	\Delta \tau^{(\mathrm{Doppler})} \nonumber \\
	-n^i\Big[\xi_i(\bar{x}^{\mu}_B(\tau_{\mathcal{E}}))&-2\xi_i(\bar{x}^{\mu}_M(\tau_{\mathcal{E}}+L))+\xi_i(\bar{x}^{\mu}_B(\tau_{\mathcal{E}}+2L))\Big] \, .
\end{align}
Next, we evaluate the Shapiro term in the new gauge. Applying the gauge transformation in Eq.~\eqref{eqn:gauge_transformation} to the null geodesics in Eq.~\eqref{eqn:null_geodesic}, the Shapiro term in Eq.~\eqref{eqn:dse} becomes
\begin{align}\label{eqn:Shapiro_gauge_2}
	\Delta \tau^{(\mathrm{Shapiro})\prime} &=	\Delta \tau^{(\mathrm{Shapiro})}+ \left[\xi_0(\bar{x}^{\mu}_B(\tau_{\mathcal{E}}))-\xi_0(\bar{x}^{\mu}_B(\tau_{\mathcal{E}}+2L))\right]  \nonumber \\
	+ n^i\Big[\xi_i(\bar{x}^{\mu}_B(\tau_{\mathcal{E}}))&-2\xi_i(\bar{x}^{\mu}_M(\tau_{\mathcal{E}}+L))+\xi_i(\bar{x}^{\mu}_B(\tau_{\mathcal{E}}+2L))\Big] \, .
\end{align}
Here we have used the fundamental theorem of calculus, written as $\int_{\lambda_1}^{\lambda_2}d\lambda\, n^{\mu}_{\pm}\partial_{\mu}\xi_{\nu}\big|_{\bar{x}^{\mu}_{\gamma^{\pm}}(\lambda)} = \xi_{\nu}(x^{\mu}_{\gamma^{\pm}}(\lambda_2))-\xi_{\nu}(\bar{x}^{\mu}_{\gamma^{\pm}}(\lambda_1))$. Finally, the Einstein term in the new gauge is obtained simply by putting Eq.~\eqref{eqn:gauge_transformation} and Eq.~\eqref{eqn:timelike_geodesic_time} into Eq.~\eqref{eqn:dse}
\begin{equation}\label{eqn:Einstein_gauge}
	\Delta \tau^{(\mathrm{Einstein})\prime} = \Delta \tau^{(\mathrm{Einstein})} - \left[\xi_0(\bar{x}^{\mu}_B(\tau_{\mathcal{E}})) - \xi_0(\bar{x}^{\mu}_B(\tau_{\mathcal{E}}+2L))\right] \, .
\end{equation}
Summing the three contributions in Eq.~\eqref{eqn:Doppler_gauge}, Eq.~\eqref{eqn:Shapiro_gauge_2} and Eq.~\eqref{eqn:Einstein_gauge}, we clearly see that the extra terms arising from the gauge transformation in Eq.~\eqref{eqn:gauge_transformation} cancel out each other, i.e. $\Delta\tau'=\Delta\tau$,  concluding the proof. This analysis highlights the importance of specifying the initial conditions and appropriately transforming them under a diffeomorphism, although the choice of initial conditions cannot affect the Fourier transformed strain in Eq.~\eqref{eqn:strain} for nonzero frequencies.%

\section{Discussions}
%%%%%%%%%%%%%%%%%%%%%%%%%%%%%%%%%%%%%%%%%%%%%%%%%
%
%       Discussions
%
%%%%%%%%%%%%%%%%%%%%%%%%%%%%%%%%%%%%%%%%%%%%%%%%%%
In this paper, we have introduced the proper time observable as a physical measure for a general gravitational perturbation in interferometer experiments. We provide a straightforward method for computing this observable and demonstrate its equivalence to the detector strain for GWs. Our approach can be applied to search for signals from BSM physics, including dark matter and vacuum fluctuations of quantum gravity.

We have so far only considered simple Michelson interferometers. The signal in more advanced interferometer setups can, in general, be derived similarly once the relevant proper time quantity is identified. Additionally, if the interferometer observable factorizes into a product of a detector response function and the metric perturbation in Fourier space, one can estimate the detector's sensitivity to a general metric perturbation by re-scaling its response in a simple Michelson setup by its sensitivity to GWs, with the signal-to-noise ratio (SNR)
\begin{align}\label{eqn:SNR}
	\mathrm{SNR}^2 \sim \int_0^{\infty}df\,\frac{1}{S_{n}^{\langle\mathrm{GW}\rangle}(f)}\frac{4|\tilde{h}(f)|^2}{\sinc^2(2\pi f L)} \, ,
\end{align}
where $S_{n}^{\langle\mathrm{GW}\rangle}(f)$ is the noise power spectral density (PSD) of the detector, commonly reported in units of Hz$^{-1}$ for the purpose of GW detection~\cite{Moore:2014lga}, $\tilde{h}(f)$ is the signal spectrum defined in Eq.~\eqref{eqn:strain}, evaluated for a simple Michelson interferometer, and the re-scaling factor of $\sinc^2(2\pi f L)$ corresponds to the response function for a simple Michelson interferometer as evident in Eq.~\eqref{eqn:strain_GW}. As an example, LIGO utilizes Fabry-P\'{e}rot cavities to enhance its sensitivity to GWs~\cite{LIGOScientific:2007fwp}. Since the reported noise PSD has already accounted for the enhancement factor, the SNR for GWs is simply given by $\mathrm{SNR}^2=\int_0^{\infty}df\,4|\tilde{h}^{(\mathrm{GW})}(f)|^2/S_{n}^{\langle\mathrm{GW}\rangle}(f)$, in agreement with Eq.~\eqref{eqn:SNR}.

In this work, we have only considered proper time observables for laser interferometers. In another work~\cite{Badurina:2024rpp}, we show that in a gradiometer setup of atom interferometers, the gauge-invariant observable defined in Ref.~\cite{Dimopoulos:2008hx} admits an analogous decomposition as in Eq.~\eqref{eqn:dse}. We suspect that this is generally true for other types of GW experiments as well, such as optically levitated sensors~\cite{Arvanitaki:2012cn, Aggarwal:2020umq}, and pulsar timing arrays~\cite{Taylor:2021yjx, NANOGrav:2023hvm}. Another possible future direction is to explore the connection between the interferometer proper time observable and correlators of worldline proper length~\cite{Sivaramakrishnan:2024ydy}.

\section{Acknowledgments}
We thank Leonardo Badurina, Yanbei Chen, Valerie Domcke, Yannis Georis, Dongjun Li, Lee McCuller, Allic Sivaramakrishnan and Jordan Wilson-Gerow for helpful discussions. We are supported by the Heising-Simons Foundation ``Observational Signatures of Quantum Gravity'' collaboration grant 2021-2817. In addition, VL recognizes support by the Network for Neutrinos, Nuclear Astrophysics and Symmetries (N3AS), through the National Science Foundation Physics Frontier Center Award \#2020275, and the Heising Simons Foundation. The work of KZ is supported by a Simons Investigator award and the U.S. Department of Energy, Office of Science, Office of High Energy Physics, under Award No. DE-SC0011632.

\bibliography{bibliography}

\end{document}